\documentclass[twocolumn]{jpsj3}

\usepackage{color}
\usepackage{times}
\usepackage{graphicx}

\title{Quantum Critical Point between Two-Channel Kondo and Fermi-Liquid Phases}

\author{Takashi Hotta}

\inst{Department of Physics, Tokyo Metropolitan University,
Hachioji, Tokyo 192-0397, Japan}

\recdate{\today}

\abst{
We investigate the emergence of quantum critical points
near a two-channel Kondo phase by evaluating an $f$-electron entropy
of a seven-orbital impurity Anderson model
hybridized with three ($\Gamma_7$ and $\Gamma_8$) conduction bands
with the use of a numerical renormalization group method.
First we consider the case of Pr$^{3+}$ ion,
in which quadrupole two-channel Kondo effect is known to occur
for the local $\Gamma_3$ doublet state.
When we control crystalline electric field (CEF) potentials so as to change the local
CEF ground state from $\Gamma_3$ doublet to $\Gamma_1$ singlet or
$\Gamma_5$ triplet, we commonly observe a residual entropy of $\log \phi$
with the golden ratio $\phi=(1+\sqrt{5})/2$,
which is equal to that for three-channel Kondo effect.
This peculiar residual entropy is also observed for the case of Nd$^{3+}$ ion,
in which magnetic two-channel Kondo phase is found to occur
for the local $\Gamma_6$ doublet state.
We envisage a scenario that the quantum critical point characterized by $\log \phi$
generally appears between two-channel Kondo and Fermi-liquid phases.
}

\begin{document}
\maketitle

\section{Introduction}

In the modern condensed matter physics,
quantum critical phenomena have attracted continuous attention,
since exotic and intriguing electronic phases have incessantly emerged
near a quantum critical point (QCP).
One of typical phenomena emerging from QCP is a non-Fermi-liquid state.
In particular, a stage for the non-Fermi-liquid ground state
is provided by two-channel Kondo effect.
After the understanding of the conventional Kondo effect,\cite{Kondo}
Coqblin and Schrieffer have derived exchange interactions
from the multiorbital Anderson model.\cite{Coqblin}
Then, a concept of multi-channel Kondo effect has been developed
on the basis of such exchange interactions,\cite{Nozieres}
as a potential source of non-Fermi-liquid phenomena.

Concerning the reality of two-channel Kondo effect,
Cox has pointed out that it actually occurs in a cubic uranium compound
with non-Kramers doublet ground state.\cite{Cox1,Cox2}
Namely, for the $\Gamma_3$ ground state,
there exist anti-ferro exchange interactions
in terms of quadrupole degree of freedom
between localized and conduction electrons,
whereas spin degree of freedom plays a role of channel.
On the other hand, a non-Fermi-liquid state has been pointed out
in a two-impurity Kondo system.\cite{Jones1,Jones2,Jones3}
This phenomenon has attracted much attention from a viewpoint of QCP
and it has been confirmed that the QCP appears at the transition
between the screened Kondo and local singlet phases.
The two-channel Kondo effect and properties of the QCP have been
vigorously discussed for a long time by numerous authors.
\cite{Koga1,Koga2,Kusunose1,Kusunose2,OSakai,Koga3,Koga4,Miyake1,
Fabrizio1,Fabrizio2,Miyake2,Fabrizio3,Koga5,Miyake3,Miyake4,Sela,
Shiina1,Hotta1,Shiina2,Hotta2,Koga6,Hotta3}

We believe that the two-channel Kondo effect itself gives an inexhaustible
source of non-trivial intriguing phenomena,
but here we turn our attention to the QCP near the two-channel Kondo phase.
In particular, we are much interested in the emergence of QCP
when two $4f$ electrons of Pr$^{3+}$ ion are hybridized with
three ($\Gamma_7$ and $\Gamma_8$) conduction bands.
The QCP in the vicinity of quadrupole two-channel Kondo phase
is expected to be experimentally observed in Pr compounds,
since there recently have been significant advances
to grasp the signal of non-Fermi-liquid behavior in Pr compounds.
\cite{review1-2-20}

In this study, we analyze a seven-orbital impurity Anderson model
hybridized with $\Gamma_7$ and $\Gamma_8$ conduction bands
by using a numerical renormalization group (NRG) method.\cite{Wilson,NRG}
For Pr$^{3+}$ ion, by controlling crystalline electric field (CEF) potentials
between two-channel Kondo and CEF singlet phases,
we find a residual entropy of $\log \phi$ with $\phi= (1+\sqrt{5})/2$,
which is the same as that for three-channel Kondo effect.
This entropy also appears between two-channel Kondo and
screened Kondo singlet phases, and
we also find it when the hybridization is increased
in the two-channel Kondo phase.
Furthermore, when we analyze the model for Nd$^{3+}$ ion
to consider magnetic two-channel Kondo phase,\cite{Hotta1}
we find a $\log \phi$ plateau in the temperature dependence of the entropy.
Then, we envisage that in general, the QCP characterized by $\log \phi$
appears between two-channel Kondo and Fermi-liquid phases.

The paper is organized as follows.
In Sect.~2, for the description of the local $f$-electron states,
first we explain the local Hamiltonian
including spin-orbit couping, CEF potentials, and Coulomb interactions
among $f$ electrons.
After checking the local $f$-electron states for Pr$^{3+}$ and Nd$^{3+}$ ions,
we construct an impurity Anderson model by considering
the hybridization between localized and conduction electrons
in $\Gamma_8$ and $\Gamma_7$ orbitals.
We also explain the NRG method to analyze the impurity Anderson model.
In Sect.~3, we review the previous results for the two-band case,
in which we have included only the hybridization in the $\Gamma_8$ electrons
for Pr$^{3+}$ and Nd$^{3+}$ ions.
In particular, we explain the two-channel Kondo effect
on the basis of a $j$-$j$ coupling scheme.
In Sect.~4, we show our NRG results for the case in which we consider
the hybridization for $\Gamma_8$ and $\Gamma_7$ electrons.
We remark the emergence of QCP's between two-channel Kondo
and Fermi-liquid phases for Pr$^{3+}$ and Nd$^{3+}$ ions.
Finally, in Sect.~5, we provide a few comments on the future problems
and a possibility to detect the present QCP in actual materials.
Throughout this paper, we use such units as $\hbar=k_{\rm B}=1$.

\section{Model and Method}

\subsection{Local Hamiltonian}

Let us start our discussion on the description of the local $f$-electron state.
For the purpose, first we consider one $f$-electron state,
which is the eigenstate of spin-orbit and CEF terms.
Then, we include Coulomb interactions among $f$ electrons.
As shown in the left part of Fig.~1, under the cubic CEF potentials,
we obtain $\Gamma_7$ doublet and $\Gamma_8$ quartet from $j=5/2$ sextet,
whereas $\Gamma_6$ doublet, $\Gamma_7$ doublet,
and $\Gamma_8$ quartet from $j=7/2$ octet.
By using those one-electron states as bases,
we express the local $f$-electron Hamiltonian as
\begin{equation}
\label{Hloc}
\begin{split}
  H_{\rm loc} &=\sum_{j, \mu, \tau} (\lambda_j  + B_{j,\mu})
  f_{j \mu \tau}^{\dag} f_{j \mu \tau} + n E_{f} \\
  &+\sum_{j_1\sim  j_4} \sum_{\mu_1 \sim \mu_4}
  \sum_{\tau_1 \sim \tau_4} 
  I^{j_1 j_2, j_3 j_4}_{\mu_1 \tau_1 \mu_2 \tau_2, \mu_3 \tau_3 \mu_4 \tau_4} \\
 &\times  f_{j_1 \mu_1 \tau_1}^{\dag} f_{j_2 \mu_2 \tau_2}^{\dag}
  f_{j_3 \mu_3 \tau_3}  f_{j_4 \mu_4 \tau_4},
\end{split}
\end{equation}
where $f_{j \mu\tau}$ denotes the annihilation operator of
a localized $f$ electron in the bases of $(j, \mu, \tau)$,
$j$ is the total angular momentum,
$j=5/2$ and $7/2$ are denoted by ``$a$'' and ``$b$'', respectively,
$\mu$ distinguishes the cubic irreducible representation,
$\Gamma_8$ states are distinguished by $\mu=\alpha$ and $\beta$,
while $\Gamma_7$ and $\Gamma_6$ states are labeled
by $\mu=\gamma$ and $\delta$, respectively,
$\tau$ is the pseudo-spin which distinguishes the degeneracy
concerning the time-reversal symmetry,
$n$ is the local $f$-electron number at an impurity site,
and $E_f$ is the $f$-electron level to control $n$.
Throughout this paper, the energy unit is set as eV.

\begin{figure}[t]
\centering
\includegraphics[width=8truecm]{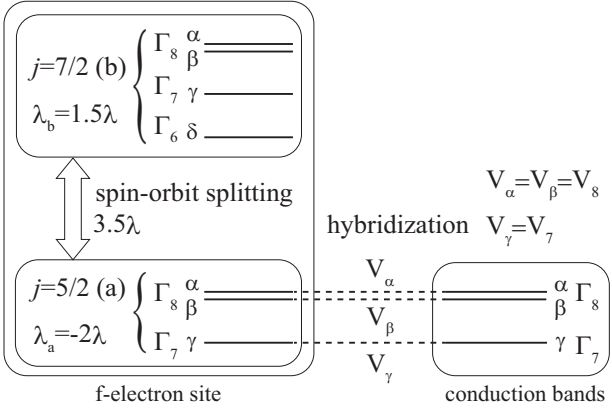}
\caption{
Schematic view of the seven-orbital impurity Anderson model.
The left part of this figure denotes the local $f$-electron states
described by a $j$-$j$ coupling scheme,
in which Coulomb interactions among such states are further included.
The lower right part indicates three conduction electron bands,
hybridized with the $j=5/2$ states of the same irreducible representations.
}
\end{figure}

As for the spin-orbit term, we obtain
\begin{equation}
\lambda_a=-2\lambda,~\lambda_b=(3/2)\lambda,
\end{equation}
where $\lambda$ is the spin-orbit coupling of $f$ electron.
In this paper, we set $\lambda=0.1$ and $0.11$
for Pr and Nd ions, respectively.
Concerning the CEF potential term for $j=5/2$,
we obtain
\begin{equation}
\begin{split}
B_{a,\alpha} & =B_{a,\beta}=1320 B_4^0/7,\\
B_{a,\gamma} &=-2640 B_4^0/7,
\end{split}
\end{equation}
where $B_4^0$ denotes the fourth-order CEF parameter
in the table of Hutchings for the angular momentum $\ell=3$.
\cite{Hutchings}
Note that the sixth-order CEF potential term $B_6^0$
does not appear for $j=5/2$, since the maximum size
of the change of the total angular momentum is less than six
in this case.
On the other hand, for $j=7/2$, we obtain
\begin{equation}
\begin{split}
B_{b,\alpha} &=B_{b,\beta}=360B_4^0/7+2880B_6^0,\\
B_{b,\gamma} &=-3240B_4^0/7-2160 B_6^0,\\
B_{b,\delta} &=360 B_4^0-3600 B_6^0/7.
\end{split}
\end{equation}
Note that the $B_6^0$ terms turn to appear in this case.
In the present calculations, we treat $B_4^0$ and $B_6^0$
as parameters.

The Coulomb interaction $I$ is expressed as
\begin{equation}
\begin{split}
&I^{j_1 j_2, j_3 j_4}_{\mu_1 \tau_1 \mu_2 \tau_2, \mu_3 \tau_3 \mu_4 \tau_4}
= \sum_{m_1\sim  m_4} \sum_{\sigma, \sigma'}
C_{j_1 \mu_1 \tau_1,m_1 \sigma} \\
&\times C_{j_2 \mu_2 \tau_2,m_2 \sigma'}
C_{j_3 \mu_3 \tau_3,m_3 \sigma'}
C_{j_4 \mu_4 \tau_4,m_4 \sigma}\\
&\times \sum_{k=0}^{6} F^k c_k(m_1,m_4)c_k(m_2,m_3),
\end{split}
\end{equation}
where $m$ is the $z$ component of the angular momentum $\ell=3$,
$\sigma$ denotes a real spin,
the sum of $k$ is limited by the Wigner-Eckart theorem to the even numbers,
$F^k$ indicates the Slater-Condon parameter,
$c_k$ is the Gaunt coefficient,\cite{Slater}
and $C$ denotes the coefficient in the transformation of
\begin{equation}
 f_{j \mu \tau}=\sum_{m,\sigma}C_{j \mu \tau, m \sigma}f_{m \sigma}.
\end{equation}
Although the Slater-Condon parameters of the material should be determined
from experimental results, here we simply set the ratio as~\cite{Hotta-Harima}
\begin{equation}
F^0/10=F^2/5=F^4/3=F^6=U,
\end{equation}
where $U$ is the Hund's rule interaction among $f$ orbitals.
In this study, we set $U$ as 1 eV.

\subsection{Local $f$-electron states}

First let us consider the local CEF ground-state phase diagram
for the case of $n=2$.
The ground-state multiplet for $B_4^0=B_6^0=0$ is
characterized by total angular momentum $J=4$.
Under the cubic CEF potentials, the nonet of $J=4$ is split into
four groups as $\Gamma_1$ singlet, $\Gamma_3$ doublet,
$\Gamma_4$ triplet, and $\Gamma_5$ triplet.
Among them, $\Gamma_4$ triplet does not appear as
a solo ground state under the cubic CEF potential
with $O_{\rm h}$ symmetry.
Then, we obtain three local ground states for $n=2$,
as shown in Fig.~2(a).

Roughly speaking, we obtain $\Gamma_1$ singlet for $B_4^0>0$,
whereas $\Gamma_5$ triplet appears for $B_4^0<0$.
Here we recall the fact that $f^1$ local ground state is
$\Gamma_7$ and $\Gamma_8$ for $B_4^0>0$ and $B_4^0<0$,
respectively.
When we accommodate two electrons into these situations,
we easily obtain $\Gamma_1$ singlet and $\Gamma_5$ triplet
by standard positive Hund's rule interaction.
As for $\Gamma_3$ doublet state, it appears for $B_6^0>0$
near the region of $B_4^0 \approx 0$.
The stabilization of $\Gamma_3$ doublet is understood
by effective negative Hund's rule interaction,
which depends on $B_6^0$.\cite{Hotta2}

In the following calculations, we use the parametrization as
\begin{equation}
B_4^0=Wx/F(4),~~B_6^0=W(1-|x|)/F(6),
\end{equation}
where $x$ specifies the CEF scheme for the $O_{\rm h}$ point group,
whereas $W$ determines the energy scale for the CEF potentials.\cite{LLW}
We choose $F(4)=15$ and $F(6)=180$ for $\ell=3$.\cite{Hutchings}
The trajectory of $B_4^0$ and $B_6^0$ for $-1 \le x \le 1$
with a fixed value of $|W|$ ($W>0$ and $W<0$)
forms a rhombus on the $(B_4^0, B_6^0)$ plane.
The trajectories for $|W|=0.001$ and $0.002$ are shown in red and blue
rhombuses, respectively, in Figs.~2.

In Fig.~2(b), we show the local CEF ground-state phase diagram
for the case of $n=3$.
The ground-state multiplet for $B_4^0=B_6^0=0$ is
characterized by $J=9/2$.
Under the cubic CEF potentials, the dectet of $J=9/2$ is split into
three groups as one $\Gamma_6$ doublet and two $\Gamma_8$ quartets.
Then, we obtain two local ground states for $n=3$,
as shown in Fig.~2(b).
Note that the $\Gamma_6$ doublet appears in the region of $B_6^0>0$,
which is the same side as that of the $\Gamma_3$ doublet.
When we consider the $f$-electron states on the basis of a $j$-$j$
coupling scheme, the $\Gamma_6$ doublet for $n=3$ is obtained
by putting one electron to the vacant orbital in the $\Gamma_3$
doublet for $n=2$.\cite{Hotta1,Kubo}
This point will be discussed later again.

\begin{figure}[t]
\centering
\includegraphics[width=8truecm]{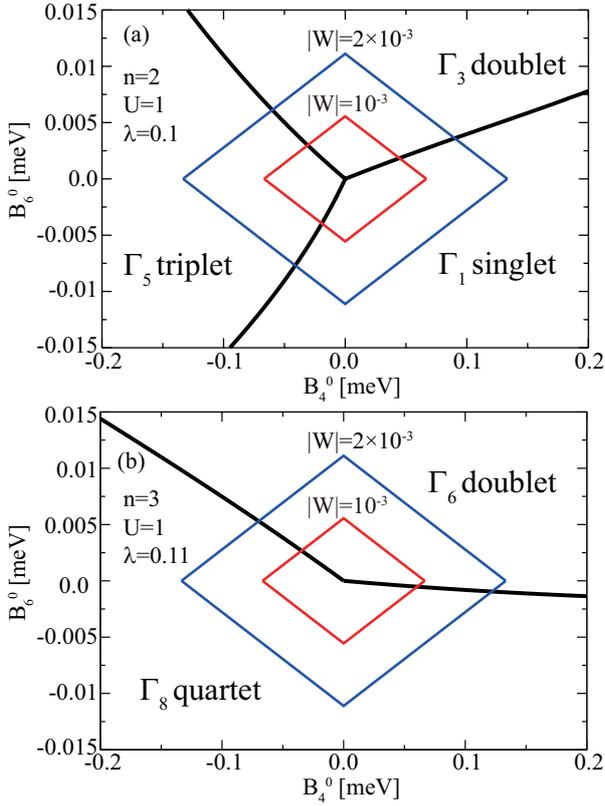}
\caption{(Color online) Local CEF ground-state phase diagrams
on the $(B_4^0, B_6^0)$ plane for (a) $n=2$ with $\lambda=0.1$
and (b) $n=3$ with $\lambda=0.11$.
Note that we set $U=1$ for both cases.
Red and blue rhombuses denote the trajectories of
$B_4^0=Wx/15$ and $B_6^0=W(1-|x|)/180$
in the range of $-1 \le x \le 1$ for $W=\pm 0.001$ and $\pm 0.002$, respectively.
}
\end{figure}

\subsection{Seven-orbital impurity Anderson model}

To construct the impurity Anderson model,
let us include the $\Gamma_7$ and $\Gamma_8$ conduction electron bands
hybridized with localized electrons.
Since we consider the cases of $n=2$ and $3$, the local $f$-electron states
are mainly formed by $j=5/2$ electrons and the chemical potential
is situated among the $j=5/2$ sextet.
Thus, we consider only the hybridization between conduction and
$j=5/2$ electrons, as shown in Fig.~1.
Then, the seven-orbital impurity Anderson model is given by
\begin{equation}
H \!=\! \sum_{\mib{k},\mu,\tau} \varepsilon_{\mib{k}}
c_{\mib{k}\mu\tau}^{\dag} c_{\mib{k}\mu\tau}
+ \! \sum_{\mib{k},\mu,\tau} V_{\mu}
(c_{\mib{k}\mu\tau}^{\dag}f_{a\mu\tau}+{\rm h.c.})
+ \! H_{\rm loc},
\end{equation}
where $\varepsilon_{\mib{k}}$ is the dispersion of conduction electron
with wave vector $\mib{k}$,
$c_{\mib{k}\mu\tau}$ denotes an annihilation operator of
conduction electron,
and $V_{\mu}$ indicates the hybridization between localized and
conduction electrons in the $\mu$ orbitals.
Note that $V_{\alpha}=V_{\beta}$ from the cubic symmetry,
whereas $V_{\gamma}$ can take a different value
from $V_{\alpha}$ and $V_{\beta}$.
Then, we define $V_{\alpha}=V_{\beta}=V_8$ and $V_{\gamma}=V_7$,
where $V_7$ ($V_8$) denotes the hybridization between
$\Gamma_7 (\Gamma_8$) conduction and localized electrons.

In our previous research, we have discussed the case of $V_7=0$,
namely, the two-band model.
We have analyzed the seven-orbital impurity Anderson model
hybridized with $\Gamma_8$ conduction bands.
\cite{Hotta1,Hotta2,Hotta3}
Later we will review the two-band results
for the cases of $n=2$, $3$, and $4$.
After that, we will show our main results of this paper
for the three-band case with $V_7=V_8$.
In this situation, we define $V_{7}=V_{8}=V$.
We will also discuss the results for the case of $V_7 \ne V_8$.

\subsection{Method}

In this study, we analyze the model by employing the NRG method.\cite{Wilson,NRG}
In this technique, we logarithmically discretize the momentum space
to include efficiently conduction electron states near the Fermi energy.
Then, the conduction electron states are characterized by ``shells'' labeled by $N$,
and the shell of $N=0$ denotes an impurity site described by $H_{\rm loc}$.
After some algebraic calculations,
the impurity Anderson Hamiltonian is transformed into the recursive form as
\begin{equation}
H_{N+1} = \sqrt{\Lambda} H_N + \xi_N \sum_{\mu,\tau}
(c_{N \mu \tau}^{\dag}c_{N+1 \mu \tau}+{\rm h.c.}),
\end{equation}
where $\Lambda$ indicates a parameter to control the logarithmic discretization,
$c_{N \tau \sigma}$ denotes the annihilation operator of the conduction electron
in the $N$-shell, and $\xi_N$ indicates the ``hopping'' of the electron between
$N$- and $(N+1)$-shells, expressed by
\begin{equation}
\xi_N=\frac{(1+\Lambda^{-1})(1-\Lambda^{-N-1})}
{2\sqrt{(1-\Lambda^{-2N-1})(1-\Lambda^{-2N-3})}}.
\end{equation}
The initial term $H_0$ is given by
\begin{equation}
H_0=\Lambda^{-1/2} [H_{\rm loc}
+\sum_{\mu,\tau} V_{\mu} (c_{0 \mu \tau}^{\dag} f_{a \mu \tau}+{\rm h.c.})].
\end{equation}

To calculate thermodynamic quantities,
we evaluate the free energy $F$ for the local $f$ electron in each step as
\begin{eqnarray}
F_N = -T (\ln {\rm Tr} e^{-H_N/T} - \ln {\rm Tr} e^{-H_N^0/T}),
\end{eqnarray}
where a temperature $T$ is defined as $T=\Lambda^{-(N-1)/2}$
at each step in the NRG calculation and $H_N^0$ indicates
the Hamiltonian without the impurity and hybridization terms.
Then, we obtain the entropy $S_{\rm imp}$ as
$S_{\rm imp}=-\partial F/\partial T$
and the specific heat $C_{\rm imp}$ is evaluated by
$C_{\rm imp}=-T\partial^2 F/\partial T^2$.

In the NRG calculation, we keep $M$ low-energy states
for each renormalization step.
In this paper, for the two-band case with $V_7=0$,
we mainly set $\Lambda=5$ and $M=2,500 \sim 4,000$,
whereas we use $\Lambda=8$ and $M=5,000$
for the three-band case with $V_7>0$ and $V_8>0$.
In the NRG calculation, the energy scale is
a half of conduction band width,
which is set as unity in the present paper.

\section{Review of the Results for the Two-Band Model}

\begin{figure}[t]
\centering
\includegraphics[width=8truecm]{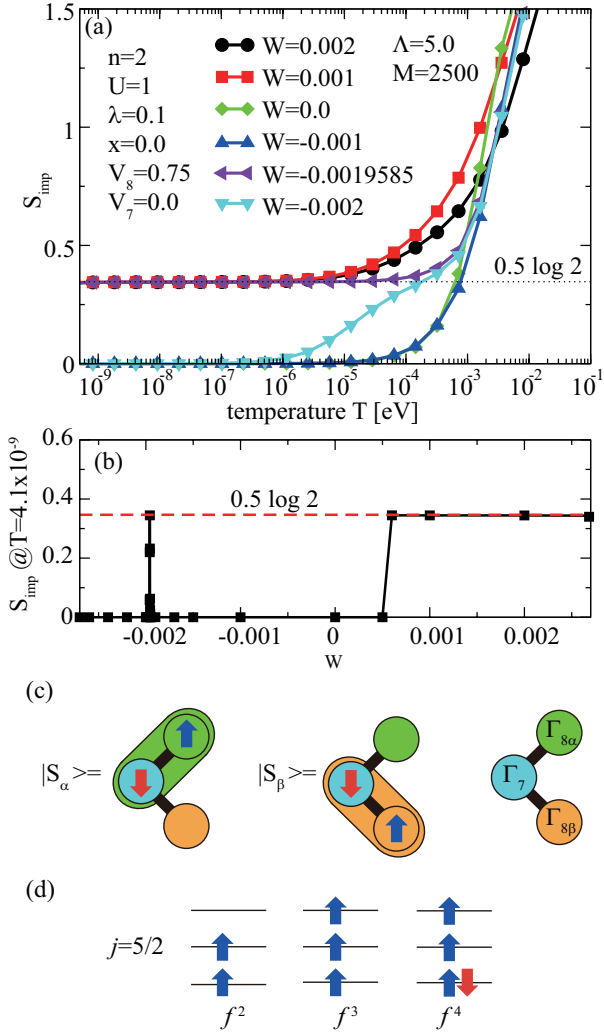}
\caption{
(Color online) (a) Entropies for $n=2$ and $V_7=0$ on the line of $B_4^0=0$.
Here we set $x=0$ and change $W$ between $W=-0.002$ and $0.002$.
(b) Residual entropies at $T=4.1 \times 10^{-9}$ vs. $W$ for $x=0$.
(c) Schematic views for the major components of the local $\Gamma_3$ states
in the $j$-$j$ coupling scheme.
Note that the oval symbolically indicates the singlet state
between $\Gamma_7$ and $\Gamma_8$ orbitals.
(d) Electron configurations in the $j$-$j$ coupling scheme for $n=2$, $3$, and $4$.
Here we accommodate electrons with pseudo-spins (blue up and red down)
in the $j=5/2$ sextet.
}
\end{figure}

Before proceedings to the discussion on the results for the three-band case,
let us review our previous results for the two-band model,
corresponding to the case for $V_7=0$ and $V_8 \ne 0$.
\cite{Hotta1,Hotta2,Hotta3}
First we consider the case of Pr$^{3+}$ ion with $n=2$.\cite{Hotta2}
In Fig.~3(a), we depict some results of $f$-electron entropies,
when we change the values of $W$ for $x=0$
from $W=0.002$ ($\Gamma_3$ doublet)
to $W=-0.002$ ($\Gamma_1$ singlet).
For $W=0.002$ and $0.001$, we observe a residual entropy of
$0.5 \log 2$ at low temperatures,
suggesting the appearance of the two-channel Kondo effect.

Here we turn our attention to Fig.~3(b),
in which we show residual entropies at $T=4.1 \times 10^{-9}$
for $-0.0027 \leq W \leq 0.0027$.
When we decrease $W$ from $W=0.0027$ to $0$ through $W=0.001$,
we observe that the entropy suddenly becomes zero
from $0.5 \log 2$ at $W \approx 5.0 \times 10^{-4}$
between $W=0.001$ and $W=0$,
suggesting the change from the two-channel Kondo to
the local Fermi-liquid phases.
Note that the local $\Gamma_3$ state is found for $W>0$,
as shown in Fig.~2(a), but due to the hybridization effect,
the singlet phase is obtained even for $W>0$.
Here we emphasize that the entropy change seems to
occur discontinuously and anomalous behavior
does not appear in the region between two-channel Kondo
and Fermi-liquid phases.

Prior to the discussion on the results in the region of $W<0$,
let us discuss the mechanism of the two-channel Kondo effect.
As mentioned in Sect.~1, Cox explained that the two-channel
Kondo effect occurs due to the exchange of $\Gamma_3$
quadrupole degrees of freedom,\cite{Cox1}
but we emphasize that it is quite useful to interpret such a picture
on the basis of a $j$-$j$ coupling scheme.\cite{Hotta2}
The local CEF states are composed of the electrons in the $j=5/2$
and $7/2$ states, but they are well approximated
by two-electron states among the $j=5/2$ sextet.
Then, the local $\Gamma_3$ states are well described by
\begin{equation}
\begin{split}
|\Gamma_{3\alpha} \rangle &\approx
\sqrt{\frac{16}{21}} |S_{\alpha} \rangle 
+ \sqrt{\frac{5}{21}} |S_8^{(1)}\rangle, \\
|\Gamma_{3\beta} \rangle &\approx
\sqrt{\frac{16}{21}}|S_{\beta} \rangle
+\sqrt{\frac{5}{21}}|S_8^{(2)}\rangle,
\end{split}
\end{equation}
where the major components,
$|S_{\alpha} \rangle$ and $|S_{\beta}\rangle$,
indicate the singlets between $\Gamma_7$ and $\Gamma_8$ orbitals,
whereas the minor components,
$|S_8^{(1)}\rangle$ and $|S_8^{(2)}\rangle$,
denote the singlets composed of two $\Gamma_8$ electrons.
As schematically shown in Fig.~3(c),
$|S_{\alpha} \rangle$ and $|S_{\beta}\rangle$ are,
respectively, given by
\begin{equation}
\label{Gamma3}
\begin{split}
|S_{\alpha} \rangle &= 
\sqrt{\frac{1}{2}}
\left( f^{\dag}_{a \gamma \uparrow} f^{\dag}_{a \alpha \downarrow}
-f^{\dag}_{a \gamma \downarrow} f^{\dag}_{a \alpha \uparrow}
\right) |0\rangle, \\
|S_{\beta} \rangle &=
\sqrt{\frac{1}{2}}
\left( f^{\dag}_{a \gamma \uparrow} f^{\dag}_{a \beta \downarrow}
-f^{\dag}_{a \gamma \downarrow} f^{\dag}_{a \beta \uparrow}
\right) |0\rangle.
\end{split}
\end{equation}
On the other hand, $|S_8^{(1)}\rangle$ and $|S_8^{(2)}\rangle$ are given by
\begin{equation}
\begin{split}
|S_8^{(1)}\rangle &=
\sqrt{\frac{1}{2}}
\left(f^{\dag}_{a \beta \uparrow} f^{\dag}_{a \beta \downarrow}
- f^{\dag}_{a \alpha \uparrow} f^{\dag}_{a \alpha \downarrow}
\right) |0\rangle,\\
|S_8^{(2)}\rangle &=
\sqrt{\frac{1}{2}}
\left( f^{\dag}_{a \alpha \uparrow} f^{\dag}_{a \beta \downarrow}
-f^{\dag}_{a \alpha \downarrow} f^{\dag}_{a \beta \uparrow}
\right) |0\rangle,
\end{split}
\end{equation}
respectively.

As mentioned above, the major components of the local $\Gamma_3$ states
are given by the singlets between $\Gamma_7$ and $\Gamma_8$ orbitals.
This picture of the composite degree of freedom is useful to promote our
understanding on the quadrupole two-channel Kondo effect.
Here we note that the $\Gamma_3$ states are classified
by the orbital degree of freedom in $\Gamma_8$,
since $\Gamma_8$ is isomorphic to the direct product of
$\Gamma_6$ and $\Gamma_3$.
When we introduce orbital operators $\mib{T}$ and $\mib{\tau}_{\sigma}$
to express the local $\Gamma_3$ state and $\Gamma_8$ conduction electron
with pseudo-spin $\sigma$, respectively,\cite{Hotta2}
we obtain the exchange term as
$J (\mib{\tau}_{\uparrow}+\mib{\tau}_{\downarrow})\cdot \mib{T}$
with the anti-ferro orbital exchange interaction $J$,
leading to the same model as that of
Nozi\'eres and Blandin.\cite{Nozieres}

Now let us return to Fig.~3(a).
When we further decrease $W$ from $W=0$ in the range of $W<0$,
we find a signal of the appearance of QCP.
Namely, both for $W=-0.001$ and $-0.002$,
the entropy is zero at low temperatures,
while at $W=-0.0019585$, we numerically observe
the residual entropy of $0.5 \log 2$ at low temperatures.
Note that this entropy immediately disappears
when we change the value of $W$ even slightly,
as we observe in the result for $W=-0.002$.
As shown in Fig.~3(b),
the entropies at $T=4.1 \times 10^{-9}$ are zeros
and we find a sharp peak at $W=-0.0019585$.
This is the well-known quantum critical behavior
emerging between CEF singlet and Kondo singlet phases.
\cite{Koga1,Koga2,Kusunose1,Kusunose2,OSakai,Koga3,Koga4,Miyake1,
Fabrizio1,Fabrizio2,Miyake2,Fabrizio3,Koga5,Miyake3,Miyake4,Sela,
Shiina1,Hotta1,Shiina2,Hotta2,Koga6,Hotta3}
Here we show only the QCP on the line of $B_4^0=0$,
but this QCP is considered to form the curve
along the boundary between $\Gamma_1$ and $\Gamma_5$
ground states in Fig.~2(a).
We note that the quantum critical curve always appears
in the $\Gamma_1$ region.
Note also that the curve seems to merge to the two-channel
Kondo phase.

Here we provide a short comment on the emergence of
two-channel Kondo effect for $n=4$.\cite{Hotta3}
It is useful to consider the $f^4$-electron configurations
in the $j$-$j$ coupling scheme.
As schematically shown in Fig.~3(d), when we accommodate
four $f$ electrons in the $j=5/2$ sextet, we find two $f$ holes there.
Namely, the electron-hole relation between $n=2$ and $n=4$
approximately holds on the basis of the $j$-$j$ coupling scheme.
Thus, we expect to observe the quadrupole two-channel Kondo effect
even for $n=4$, corresponding to Np$^{3+}$ and Pu$^{4+}$ ions.\cite{Hotta3}

\begin{figure}[t]
\centering
\includegraphics[width=8truecm]{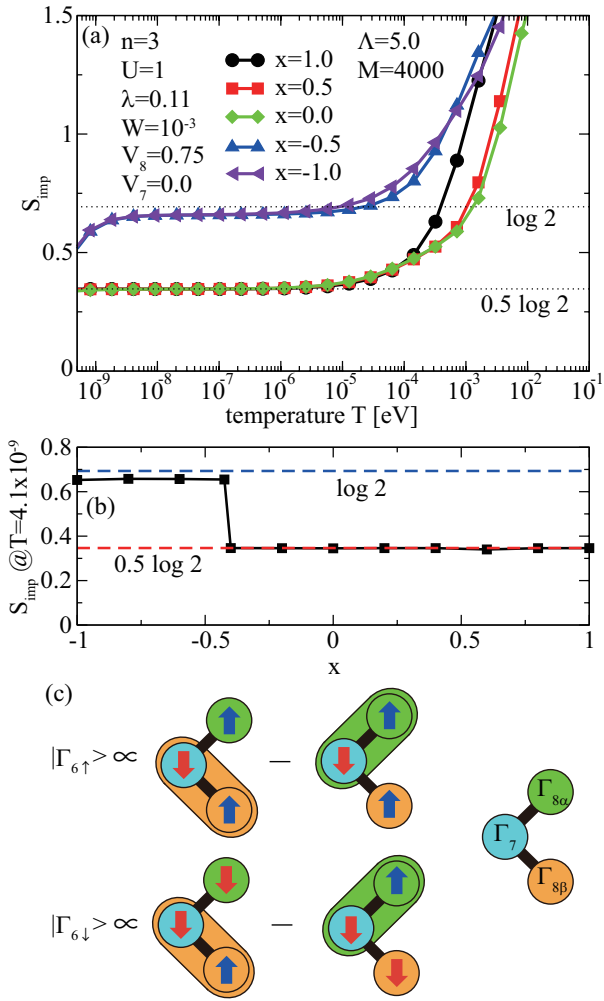}
\caption{(Color online) (a) Entropies for $n=3$ and $V_7=0$
on the lines of $W=0.001$ and $-1 \leq x \leq 1$.
(b) Residual entropies at $T=4.1 \times 10^{-9}$ vs. $x$ for $W=0.001$.
(c) Schematic views for the main components of the local $\Gamma_6$ states
in the $j$-$j$ coupling scheme.
}
\end{figure}

Now we consider the case of Nd$^{3+}$ ion with $n=3$.\cite{Hotta1}
In Fig.~4(a), we depict some results of $f$-electron entropies,
when we change the values of $x$ for $W=0.001$
from $x=1$ ($\Gamma_6$ doublet)
to $x=-1$ ($\Gamma_8$ quartet).
For $x=1$, $0.5$, and $0$, we observe residual entropies of
$0.5 \log 2$ at low temperatures,
indicating the two-channel Kondo effect.
For $x=-0.5$ and $x=-1$,
we find entropy plateaus with the value near $\log 2$,
but they are eventually released for $T<10^{-9}$.

In Fig.~4(b), we show residual entropies
at $T=4.1 \times 10^{-9}$ between $-1 \leq x \leq 1$.
For $-0.4 \leq x \leq 1$, we observe the residual entropy of
$0.5 \log 2$, suggesting the appearance of two-channel Kondo effect.
At $x \approx 0.4$, the residual entropy of $0.5 \log 2$
seems to change suddenly to the value near $\log 2$.
Roughly speaking, the region with residual entropy $\log 2$
corresponds to that of $\Gamma_8$ quartet in Fig.~2(b).
The $\Gamma_8$ quartet is approximately obtained
by the addition of one $\Gamma_8$ electron to the double
occupied $\Gamma_7$ states on the basis of the $j$-$j$ coupling scheme.
When we include the hybridization with $\Gamma_8$ conduction
electrons, we understand the appearance of the Kondo effect,
since the $\Gamma_7$ degree of freedom is suppressed.
Thus, the residual entropy near $\log 2$ in Fig.~4(b) should be
eventually released when we decrease the temperature,
leading to the Fermi-liquid state.

To understand the two-channel Kondo effect
emerging from the $\Gamma_6$ doublet,
it is again useful to consider the $\Gamma_6$ states
on the basis of the $j$-$j$ coupling scheme.
After some algebraic calculations,
the major components of the $\Gamma_6$ states are found to
be expressed by three pseudo-spins on
$\Gamma_7$ and $\Gamma_8$ orbitals,
as shown in Fig.~4(c).
Namely, we obtain~\cite{Hotta1,Kubo}
\begin{equation}
\begin{split}
|\Gamma_6, \uparrow \rangle &= \sqrt{\frac{1}{3}}
\left( f^{\dag}_{a \alpha \uparrow} |S_{\beta}\rangle
-f^{\dag}_{a \beta \uparrow} |S_{\alpha}\rangle \right),\\
|\Gamma_6, \downarrow \rangle &=\sqrt{\frac{1}{3}}
\left( f^{\dag}_{a \alpha \downarrow} |S_{\beta}\rangle
-f^{\dag}_{a \beta \downarrow} |S_{\alpha}\rangle \right).
\end{split}
\end{equation}
As discussed above, the main components of the $\Gamma_3$ doublet
states for $n=2$ are expressed by $|S_{\alpha}\rangle$ and
$|S_{\beta}\rangle$.
Then, we obtain the $\Gamma_6$ doublet states
for $n=3$ by adding one $\Gamma_8$ electron to
the $\Gamma_3$ states for $n=2$,
as shown in Fig.~4(c).

As emphasized in Ref.~\citen{Hotta1},
on the basis of the local $\Gamma_6$ states composed of 
three pseudo-spins,
we envisage a picture that the local $\Gamma_7$ pseudo-spin
is screened by $\Gamma_8$ electrons,
when we include the hybridization between localized and
conduction $\Gamma_8$ electrons.
The present picture has been actually explained by 
the extended $s$-$d$ model.\cite{Hotta1}
We believe that it is the realization of the magnetic two-channel
Kondo effect, originally raised by Nozi\'eres and Blandin.\cite{Nozieres}

\section{Calculation Results for the Three-Band Model}

\subsection{Results for $n=2$: Effect of CEF potentials}

Now we show our present results for the three-band case
with $V_7=V_8=V$.
In this subsection, let us discuss the effect of CEF potentials
on the emergence of QCP for the case of $n=2$.
In Fig.~5(a), we pick up some results when we change the CEF parameters
from $x=0$ ($\Gamma_3$ doublet) to $x=1$ ($\Gamma_1$ singlet)
for $W=0.001$.
In Fig.~5(b), we plot $S_{\rm imp}$ at $T=7.5 \times 10^{-9}$
as a function of $x$.

For $x=0$ and $0.17$, we find a residual entropy of $0.5 \log 2$
at low temperatures,
suggesting the appearance of the two-channel Kondo effect,
even when we consider the hybridization with $\Gamma_7$ band
in addition to those with $\Gamma_8$ bands.
It is useful to consider the local $\Gamma_3$ states
on the basis of the $j$-$j$ coupling scheme, as shown in Fig.~3(c).
Namely, the doublet is expressed by the composite states,
which are two singlets between
$\Gamma_7$ and $\Gamma_8$ orbitals.
When the hybridization occurs only between $\Gamma_8$orbitals,
there appear anti-ferro exchange interactions
in terms of orbital degree of freedom
between localized and conduction $\Gamma_8$ electrons,
whereas pseudo-spin plays a role of channel.
In this case, the localized $\Gamma_7$ orbital plays no 
role in the hybridization process, after the formation
of the singlet between $\Gamma_7$ and $\Gamma_8$ electrons.

When we further include the hybridization of $\Gamma_7$,
there appears a process in which the singlets composed of
$\Gamma_3$ doublets are destroyed by the formation of
another singlet between localized and conduction $\Gamma_7$ electrons.
Thus, we envision the competition between $\Gamma_7-\Gamma_7$
and $\Gamma_7-\Gamma_8$ singlets, leading to a QCP.
For the three-band case, there occurs three channels,
in which two of them are $\Gamma_8$ pseudo-spins
and another is $\Gamma_7$ orbital,
as schematically shown in Fig.~5(c).

Here we consider the case of $V_7 \gg V_8$,
in which the two-channel Kondo phase should be suppressed,
since the singlets composed of $\Gamma_3$ doublets are
considered to be destroyed by the screening of
the $\Gamma_7$ pseudo-spin.
In fact, in the limit of $V_8=0$, it is easy to imagine the appearance
of the underscreening Kondo effect concerning the $\Gamma_7$
pseudo-spin with localized $\Gamma_8$ quartet.
Thus, we deduce that there appears additional $\Gamma_7$ orbital
channel in the two-channel Kondo phase for the three-band case.

Next we turn our attention to the results for $x=0.2$ and $0.3$
in Fig.~5(a).
Although these $x$ values for $W=0.001$ are included
in the $\Gamma_3$ doublet in Fig.~1,
it seems that we arrive at the singlet phase
due to the hybridization effect,
since the residual entropy becomes zero.
It is difficult to prove the Fermi-liquid phase only by the entropy results,
but we deduce it also from the energy spectrum data.

\begin{figure}[t]
\centering
\includegraphics[width=8truecm]{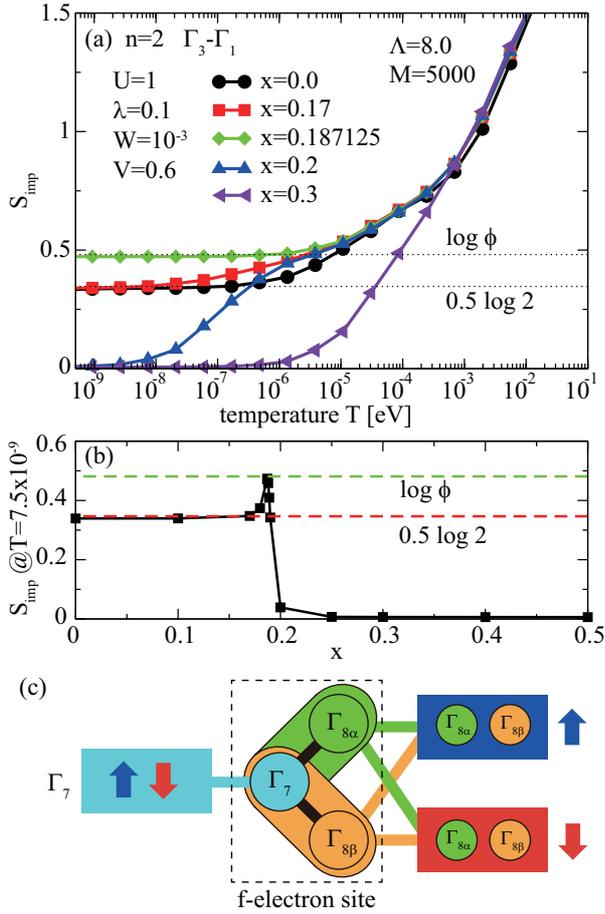}
\caption{(Color online) (a) Entropies for $n=2$, $W=0.001$, and
$0 \leq x \leq 0.3$ between $\Gamma_3$ and $\Gamma_1$.
(b) Residual entropies at $T=7.5 \times 10^{-9}$ vs. $x$ for $W=0.001$.
(c) Schematic view for the three channels when the local $\Gamma_3$
states are hybridized with three conduction bands for $n=2$.
}
\end{figure}

Now we focus on the result at $x=0.187125$,
in which we find a residual entropy, suggesting the unstable fixed point.
In fact, it disappears when we slightly change $x$.
We call it the $\Gamma_3-\Gamma_1$ QCP,
but the value of the residual entropy should be carefully discussed.
It is between $0.5\log 2$ and $\log 2$, but at first glance,
we could not understand the meaning of that value.
Here we recall that the QCP between two different Fermi-liquid phases
is characterized by the residual entropy of $0.5\log 2$,
equal to that for the two-channel Kondo effect.
Then, we hit upon a simple idea that the QCP between
two-channel Kondo and Fermi-liquid phases
is characterized by the residual entropy
equal to that for the three-channel Kondo effect.
The analytic value of the residual entropy $S_{\rm ana}$
for multi-channel Kondo effect is given by$~$\cite{Affleck}
\begin{equation}
\label{eq:Sana}
S_{\rm ana}=\log \frac{\sin [(2S +1)\pi/(n_{\rm c}+2)]}
{\sin [\pi/(n_{\rm c}+2)]},
\end{equation}
where $S$ denotes the impurity spin and
$n_{\rm c}$ indicates the channel number.
In the present case,
the $\Gamma_3$ doublet is effectively expressed by $S=1/2$
and we obtain $n_{\rm c}=3$ with two pseudo-spins and one orbital,
as shown in Fig.~5(c).
Thus, we obtain $S_{\rm ana}=\log \phi$
with the golden ratio $\phi=(1+\sqrt{5})/2$.

From the numerical results shown by green circles in Fig.~5(a),
we obtain $S_{\rm imp}=0.472$ at $T=7.5 \times 10^{-9}$,
which is deviated from the analytic value $S_{\rm ana}=0.481$.
For the two-channel Kondo phase at $x=0$,
we find $S_{\rm imp}=0.339$ at $T=7.5 \times 10^{-9}$,
which is also deviated from the analytic value
$S_{\rm ana}=0.5 \log 2=0.347$.
This deviation inevitably occurs in the NRG calculation
due to $\Lambda$ larger than unity and finite $M$.
Note that the deviation becomes large for $\Lambda>8$
when we fix $M$ as $M=5,000$.
We consider that the result at $x=0.187125$ suggests
the unstable fixed point characterized by the entropy $\log \phi$.
Note that this type of QCP was pointed out in a three-orbital impurity Anderson
model for a single C$_{60}$ molecule.\cite{Fabrizio3}

\begin{figure}[t]
\centering
\includegraphics[width=8truecm]{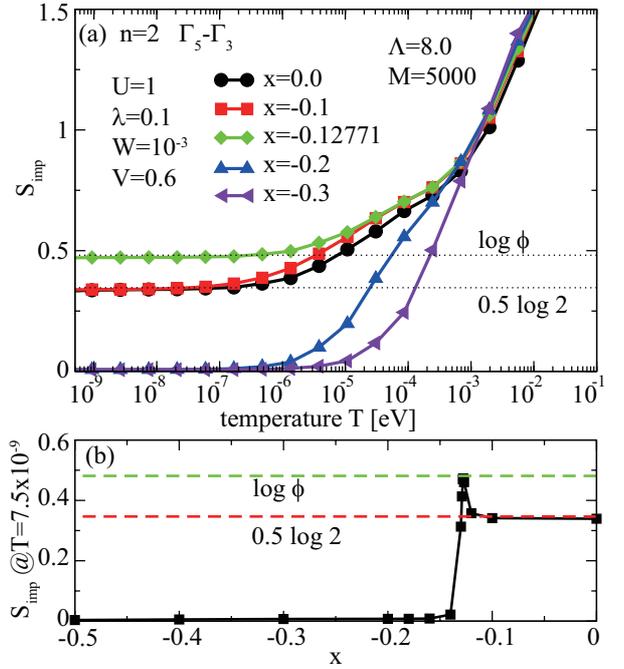}
\caption{(Color online) (a) Entropies for $n=2$, $W=0.001$, and
$-0.3 \leq x \leq 0$ between $\Gamma_3$ and $\Gamma_5$.
(b) Residual entropies at $T=7.5 \times 10^{-9}$ vs. $x$ for $W=0.001$.
}
\end{figure}

Let us now turn our attention to Fig.~5(b),
in which we show $S_{\rm imp}$ at $T=7.5 \times 10^{-9}$
as a function of $x$ to confirm the critical behavior.
We clearly find a sharp peak with the value of $\log \phi$
between the two-channel Kondo region characterized
by the entropy of $0.5 \log 2$ and the Fermi-liquid phase
characterized by zero entropy.
From these results, it is highly believed that an important aspect of
the QCP is captured, although we recognize that the existence of QCP
is not rigorously proved by the present results.

If the QCP appears between two-channel Kondo and Fermi-liquid phases,
the emergence of the QCP should {\it not} be limited to the region
between $\Gamma_3$ doublet and $\Gamma_1$ singlet.
In Fig.~6(a), we show some results when we change the CEF parameters
from $x=0$ ($\Gamma_3$ doublet) to $x=-1$ ($\Gamma_5$ triplet)
for $W=0.001$.
For $x=0$ and $-0.1$, we obtain the two-channel Kondo phase,
while for $x=-0.3$ and $-0.2$, the screened Kondo phase appears,
since in a $j$-$j$ coupling scheme, the local $\Gamma_5$ triplet is
mainly composed of two $\Gamma_8$ electrons,
which are screened by $\Gamma_8$ conduction electrons.
Between two-channel Kondo and screened Kondo phases,
we again find the residual entropy of $\log \phi$ at $x=-0.12771$.
As shown in Fig.~6(b), the entropies at $T=7.5 \times 10^{-9}$
suggest the QCP as a sharp peak,
which defines the $\Gamma_3-\Gamma_5$ QCP.
Thus, we guess that the QCP characterized by the residual entropy
of $\log \phi$ generally appears between two-channel Kodno
and Fermi-liquid phases.

\begin{figure}[t]
\centering
\includegraphics[width=8truecm]{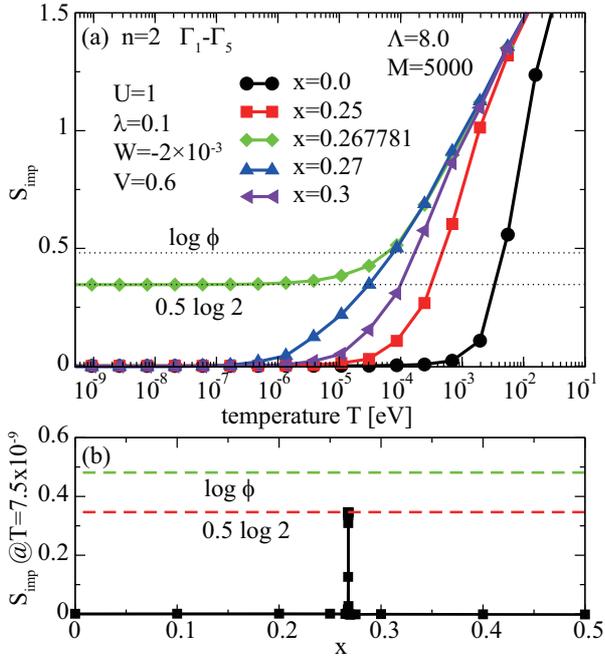}
\caption{(Color online) (a) Entropies for $n=2$, $W=-0.002$,
and $0 \leq x \leq 0.3$ between $\Gamma_5$ and $\Gamma_1$.
(b) Residual entropies at $T=7.5 \times 10^{-9}$ vs. $x$ for $W=-0.002$.
}
\end{figure}

Although we observe the $\Gamma_3-\Gamma_5$ QCP
characterized by $\log \phi$
between two-channel Kondo and screened Kondo phases,
it does not immediately mean that the property of
the $\Gamma_3-\Gamma_5$ QCP is the same
as that of the $\Gamma_3-\Gamma_1$ QCP
between two-channel Kondo and CEF singlet phases.
In particular, from a microscopic viewpoint,
we are interested in the kinds of
the three channels on the QCP.
As for the $\Gamma_3-\Gamma_1$ QCP,
we consider that two of the three channels 
are pseudo-spins and another is orbital,
as shown in Fig.~5(c) for the three-channels in the
two-channel Kondo effect emerging from
the local $\Gamma_3$ states.
Here we believe that the three channels
on the $\Gamma_3-\Gamma_5$ QCP
are the same as those on the $\Gamma_3-\Gamma_1$ QCP,
but unfortunately, we cannot prove it exactly in this study.
As mentioned above,
the local $\Gamma_5$ triplet mainly composed of
a couple of $\Gamma_8$ electrons should be
screened by $\Gamma_8$ conduction electrons.
Thus, we cannot deny a possibility that all three channels
on the $\Gamma_3-\Gamma_5$ QCP may
be related to orbital degrees of freedom.
At the present stage, it is phenomenologically shown
that the QCP characterized by $\log \phi$ appears
between two-channel Kondo and Fermi-liquid phases.
The properties of the $\Gamma_3-\Gamma_1$ and
$\Gamma_3-\Gamma_5$ QCP's are not completely clarified,
but we will mention this point later again.

Now we turn our attention to the QCP between CEF singlet and
screened Kondo singlet phases,
when we include the hybridization with three conduction bands.
In Fig.~7(a), we pick up some results when we change the CEF parameters
from $x=0$ ($\Gamma_1$ singlet) to $x=1$ ($\Gamma_5$ triplet).
Note that here we use $|W|=0.002$ to emphasize the critical behavior.
For $x=0$ and $0.25$, we find the CEF singlet phase,
while for $x=0.27$ and $0.3$, the screened Kondo singlet phase appears.
At $x=0.267781$, we observe the residual entropy of $0.5 \log 2$,
as we have expected from the QCP between two different Fermi-liquid phases.
As shown in Fig.~7(b), we observe a sharp peak with
the value of $0.5 \log 2$ around at $x \approx 0.27$.
Note again that the local $\Gamma_5$ triplet
mainly composed of two $\Gamma_8$ electrons is
screened by $\Gamma_8$ conduction electrons.
Thus, the $\Gamma_8$ conduction electrons play main roles
and the $\Gamma_7$ degree of freedom is considered to be
irrelevant on the QCP between two Fermi-liquid phases
even in the three-band case.

\begin{figure}[t]
\centering
\includegraphics[width=8truecm]{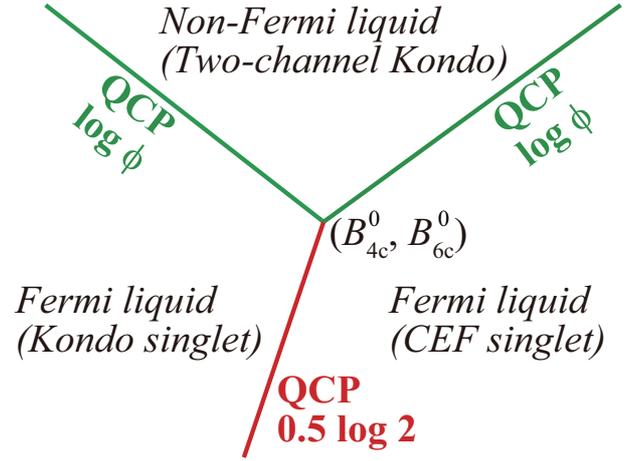}
\caption{(Color online) Schematic view for the phase diagram
on the $(B_4^0, B_6^0)$ plane,
deduced from the NRG results in Figs.~5, 6, and 7.
}
\end{figure}

To summarize this subsection,
we depict a schematic view for the phase diagram
on the $(B_4^0, B_6^0)$ plane in Fig.~8
on the basis of the results in Figs.~5, 6, and 7.
Between two-channel Kondo and Fermi-liquid
(CEF singlet or screened Kondo singlet) phases,
we deduce the appearance of the line of QCP characterized by $\log \phi$.
On the other hand, we expect another QCP line
characterized by $0.5 \log 2$ between two different Fermi-liquid phases.

We believe that the schematic phase diagram in Fig.~8 grasps
the essential points on the $(B_4^0, B_6^0)$ plane,
but we do not prove the existence
of the point $(B_{4{\rm c}}^0, B_{6{\rm c}}^0)$
at which three QCP lines are connected to one another.
Here we mention $(B_{4{\rm c}}^0, B_{6{\rm c}}^0) \ne (0, 0)$
due to the effect of hybridizations,
since the boundary curves are deviated from those among
local CEF ground states as shown in Fig.~2(a).
For instance, the region of the two-channel Kondo phase is
not exactly equal to that of the local $\Gamma_3$ state
and the QCP line characterized by $0.5 \log 2$
always appears in the region of the local $\Gamma_1$ state.

In a phenomenological level, we believe that
the kinds of the three channels on the $\Gamma_3-\Gamma_5$ QCP
are the same as those on the $\Gamma_3-\Gamma_1$ QCP,
since both are characterized by the residual entropy $\log \phi$.
It may be possible to clarify this issue when we investigate
the quantum critical behavior in the vicinity of the point of
$(B_{4{\rm c}}^0, B_{6{\rm c}}^0)$.
It is one of future problems.

\subsection{Results for $n=2$: Effect of hybridization}

Next we consider the effect of the hybridization on
the emergence of QCP by monitoring the entropies
for the fixed values of CEF parameters.
First we consider the case of $V_7=V_8=V$.
In Fig.~9(a), we show the entropies by changing $V$ from
$V=0.57$ to $V=0.7$ for $n=2$, $W=0.001$, and $x=0$.
In Fig.~9(b), we plot $S_{\rm imp}$ at $T=7.5 \times 10^{-9}$
as a function of $x$.

Now we pay our attention to the results for $V=0.57$ and $0.6$,
in which we observe the residual entropy near the value of
$0.5 \log 2$ at $T=7.5 \times 10^{-9}$,
suggesting the existence of the two-channel Kondo phase,
while for $V=0.615$ and $0.7$, we obtain the Fermi-liquid phase,
since the entropies are eventually released.
Note that in the present paper, we do not further discuss
the entropy behavior for $V$ smaller than $0.57$.
At $V=0.607841$, we again observe the QCP characterized by $\log \phi$.
This result is considered to reconfirm the emergence of QCP
induced by the hybridization
between two-channel Kondo and Fermi-liquid phases.

\begin{figure}[t]
\centering
\includegraphics[width=8truecm]{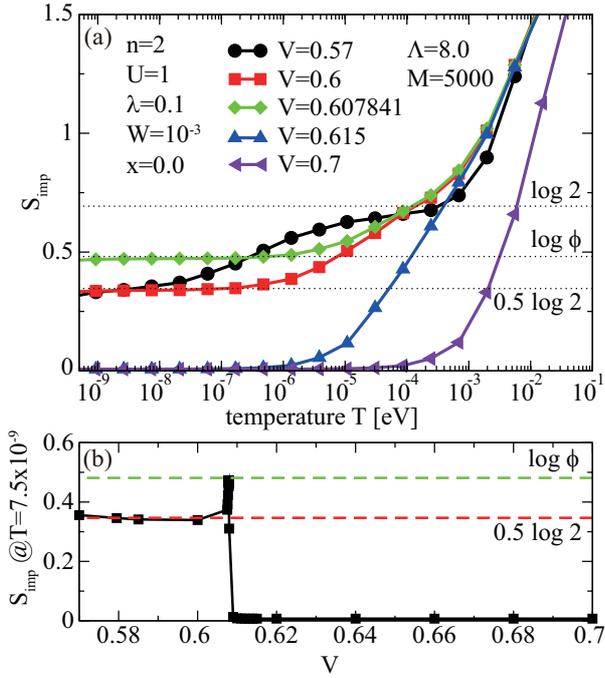}
\caption{(Color online)
(a) Entropies for $n=2$ and $0.57 \leq V \leq 0.7$ with $W=0.001$ and $x=0$.
(b) Residual entropies at $T=7.5 \times 10^{-9}$ vs. $V$ for $W=0.001$ and $x=0$.
}
\end{figure}

Next we discuss the emergence of the QCP characterized by
$\log \phi$ for the case of $V_7 \ne V_8$.
Here we shortly discuss two limiting situations,
$V_7=0$ and $V_8=0$.
The former case of $V_7=0$ corresponds to the two-band model,
as discussed in Sect.~3.
Namely, the two-channel Kondo effect has been confirmed
to appear in the region of the local $\Gamma_3$ state
for an appropriate value of $V_8$.
As shown later, when we increase the value of $V_8$
for the case of $V_7=0$,
we do not observe the QCP characterized by $\log \phi$
between two-channel Kondo and Fermi-liquid phases
in the NRG results, as shown later.

The case of $V_8=0$ apparently corresponds to
the one-band model, in which the underscreening Kondo
effect is expected to occur.
Here it is useful to recall that the local $\Gamma_3$ is expressed
by the singlets between $\Gamma_7$ and $\Gamma_8$ electrons.
For small $V_7$, the local $\Gamma_3$ doublet remains
even at low temperatures, leading to a residual entropy of $\log 2$,
when we consider the hybridization of $\Gamma_7$ electrons.
On the other hand, for large $V_7$,
the $\Gamma_7$ pseudo-spin in the local $\Gamma_3$ singlets
is screened by $\Gamma_7$ conduction electrons,
leading to a residual entropy of $\log 4$
originating from the remaining $\Gamma_8$ electron.
In any case, for the case of $V_8=0$, we do not expect
the appearance of the two-channel Kondo effect.

When we turn our attention to the case of
$V_7 \ne 0$ and $V_8 \ne 0$,
it is intuitively considered that the QCP appearing
for the case of $V_7=V_8=V$ should not disappear immediately
even when $V_7$ becomes different from $V_8$.
Furthermore, we expect that the QCP still appears
for a certain critical value of $V_8$ for the case of small $V_7$.

In Fig.~10(a), we show some results when we change the value of $V_8$
from $V_8=0.8$ to $V_8=0.85$ for $V_7=0.1$ with $W=0.001$ and $x=0$.
For $V_8=0.8$ and $0.815$, we obtain the two-channel Kondo phase,
suggested from the residual entropy of $0.5 \log 2$,
while for $V_8=0.825$ and $0.85$, the Fermi-liquid phase appears.
Between two-channel Kondo and Fermi-liquid phases,
we find the residual entropy of $\log \phi$ at $V_8=0.81936$.
As shown in Fig.~10(b), we observe the QCP as a sharp peak
in the entropies at $T=7.5 \times 10^{-9}$.

\begin{figure}[t]
\centering
\includegraphics[width=8truecm]{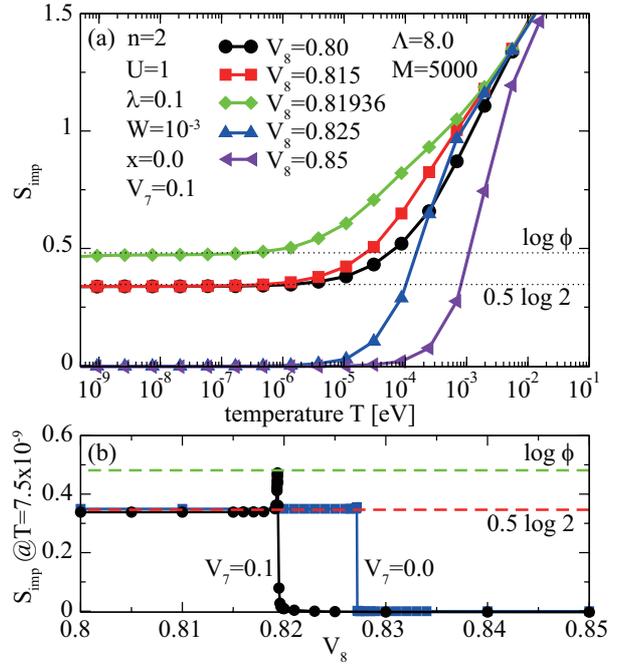}
\caption{(Color online)
(a) Entropies for $n=2$ and $0.8 \leq V_8 \leq 0.85$ for $V_7=0.1$,
$W=0.001$, and $x=0$.
(b) Residual entropies at $T=7.5 \times 10^{-9}$ vs. $V_8$ for
$V_7=0$ (blue solid square)  and $0.1$ (black solid circle).
Other parameters are the same as those in (a).
}
\end{figure}

In Fig.~10(b), we also show the entropies for the case of $V_7=0$.
Note here that we perform the NRG calculations
with $\Lambda=8$ and $M=5,000$ even for the two-band case
to keep the same conditions as those for the three-band case.
In this case, we do not observe a characteristic peak with
the value of $\log \phi$ between two-channel Kondo
and Fermi-liquid phases.
Rather we find a sudden change from $0.5 \log 2$ to zero
in the two-band case at $V_8=V_8^*=0.8272$.
This is consistent with Fig.~3(b), showing the residual entropies
when we change $W$ with $x=0$.
When we compare the results for $V_7 \ne 0$ and $V_8 \ne 0$
with those for $V_7=0$ and $V_8 \ne 0$,
we imagine that the hybridization with $\Gamma_7$
band is indispensable for the appearance of the QCP
characterized by the entropy of $\log \phi$.

\subsection{Results for $n=3$}

\begin{figure}[t]
\centering
\includegraphics[width=8truecm]{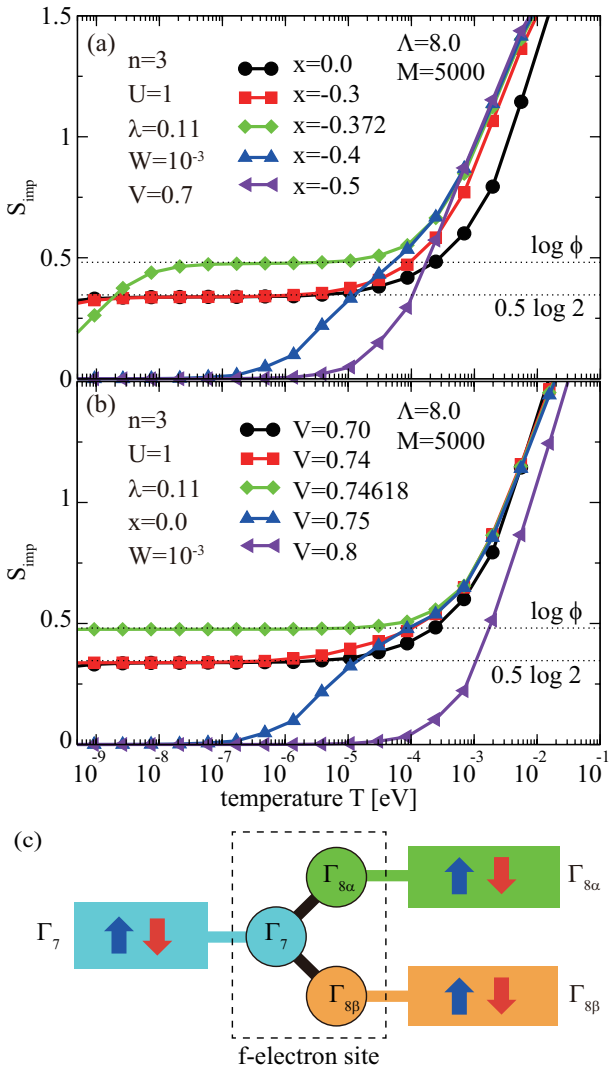}
\caption{(Color online) (a) Entropies for $W=10^{-3}$ and $-0.5 \leq x \leq 0$.
Here we set $V=0.7$.
(b)  Entropies for $0.5 \leq V \leq 0.8$. Here we fix $x=0.0$ and $W=10^{-3}$.
(c) Schematic view for the three channels when the local $\Gamma_6$
states are hybridized with three conduction bands for $n=3$.
}
\end{figure}

Thus far, we have considered the QCP
near the quadrupole two-channel Kondo phase
for the case of $n=2$.
To confirm our idea that the QCP appears between
two-channel Kondo and Fermi-liquid phases,
in this subsection, we show our NRG results for the case of $n=3$.
As mentioned in Sect.~3,
it has been shown that the magnetic two-channel Kondo effect occurs
in the $\Gamma_6$ doublet for the two-band case.\cite{Hotta1}

In Fig.~11(a),  we show some results when we change the CEF parameters
from $x=0$ ($\Gamma_6$ doublet) to $x=-1$ ($\Gamma_8$ quartet)
for $n=3$, $W=0.001$, and $V=0.7$.
For $x=0$ and $-0.3$, we obtain the magnetic two-channel Kondo phase,
while for $x=-0.4$ and $-0.5$, the screened Kondo phase appears,
since the $\Gamma_8$ quartet effectively expressed by $S=3/2$ spin
is screened by three conduction electrons.
At $x=-0.372$ between two-channel Kondo and Fermi-liquid phases,
we find an entropy plateau of $\log \phi$.
In this case, we could not observe the residual entropy of $\log \phi$
at low enough temperatures,
but the signal of QCP is considered to be obtained.

In Fig.~11(b), we show the results for $f$-electron entropies
when we change $V$ for $n=3$, $W=0.001$, and $x=0$.
For $V=0.7$ and $0.74$, we clearly observe a residual entropy
$0.5 \log 2$, suggesting the two-channel Kondo effect,
while for $V=0.75$ and $0.8$,
we find the local singlet phases, suggesting the Fermi-liquid states.
Between those two phases, at $V=0.74618$,
we again observe the residual entropy characterized by $\log \phi$.
These results for $n=3$ strongly suggest that the emergence of QCP
between magnetic two-channel Kondo and Fermi-liquid phases.

Concerning the three channels on the QCP,
it is again useful to recall the local $\Gamma_6$ doublet
on the basis of the $j$-$j$ couplings scheme,
as shown in Fig.~4(c).
As mentioned in Sect.~3, we have understood the magnetic
two-channel Kondo effect for $n=3$
by considering that the $\Gamma_7$ pseudo-spin
is screened by $\Gamma_8$ electrons.
In Fig.~11(c), we schematically show the figure
for the three channels in the two-channel Kondo
effect emerging from the local $\Gamma_6$
states for $n=3$.
When we further include the effect of hybridization of $\Gamma_7$
electron, we envisage a picture that the $\Gamma_7$ pseudo-spin
is screened by $\Gamma_8$ and $\Gamma_7$ electrons.
Namely, on the QCP characterized by $\log \phi$ for $n=3$,
impurity spin $S$ in Eq.~(\ref{eq:Sana}) denotes the
$\Gamma_7$ pseudo-spin and
all three channels ($n_{\rm c}=3$)
are given by orbital degrees of freedom.

\section{Discussion and Summary}

In this paper, we have analyzed the seven-orbital impurity Anderson
model hybridized with three ($\Gamma_7$ and $\Gamma_8$)
conduction bands with the use of the NRG method.
For Pr$^{3+}$ and Nd$^{3+}$ ions, we have evaluated
the $f$-electron entropies by controlling CEF potentials
and hybridization magnitudes.
Then, we have found the QCP characterized by the residual entropy
of $\log \phi$ between two-channel Kondo and Fermi-liquid phases.

We believe that the emergence of this QCP does not 
depend on the nature of the two-channel Kondo phase,
quadrupole or magnetic,
and also the details of the Fermi-liquid phase,
CEF or Kondo singlets.
As for the three channels the QCP,
they are considered to depend on the property of the two-channel Kondo phase.
Namely, for $n=2$, the three channels
are two pseudo-spins of $\Gamma_8$ and one orbital of $\Gamma_7$,
whereas for $n=3$, they are three orbitals of $\Gamma_8$ and $\Gamma_7$.
Note that for $n=2$, the kinds of the three channels
on the $\Gamma_3-\Gamma_5$ QCP
are the same as those on the $\Gamma_3-\Gamma_1$ QCP.
However, we cannot prove it in a microscopic level in the present study.
For the purpose, it is necessary to promote our understanding on
the phase diagram in Fig.~8 beyond the phenomenological level.
This is one of future problems.

When we depict another phase diagram concerning the hybridization, 
the QCP characterized by $\log \phi$ is considered to form a curve
on the $(V_8, V_7)$ plane.
As mentioned above, this QCP does not appear for $V_7=0$ or $V_8=0$.
For $V_8=0$, since the model is the one-band case,
we can easily understand that the QCP does not appear.
On the other hand, for the case of $V_7=0$,
the two-channel Kondo phase
is discontinuously changed to the Fermi-liquid phase
at a certain value of $V_8^*$.
The QCP curve for $V_8 \ne 0$ and $V_7 \ne 0$ seems to
merge into the point at $V_8=V_8^*$ and $V_7=0$,
but the fate of the QCP for the infinitesimal value of
$V_7$ is unclear.
This point will be investigated in future.

Furthermore, it is necessary to discuss the phase for small $V_7$ and $V_8$
with due care.
This is the reason why we have not discussed the entropy behavior
for small $V$ in Fig.~9.
If the Fermi-liquid phase appears at this region,
it may be different from that in the large hybridization region.
In such a case, we expect another QCP curve characterized by $\log \phi$
depending on the CEF parameters.
On the other hand, there is a possibility that the Fermi-liquid phase
for large $V$ is connected to that for small $V$
when we consider the region of $V_8 \ne V_7$.
It is also a future problem to clarify the phase diagram
on the $(V_8, V_7)$ plane.

In this paper, we have phenomenologically confirmed the QCP characterized
by $\log \phi$ from the evaluation of the $f$-electron entropy,
but it should be remarked that the existence of the QCP has not been proven exactly.
To pile up the evidence, it is necessary to analyze other physical quantities.
For instance, when we define a characteristic temperature $T^*$
at which the entropy $\log \phi$ is released,
it is interesting to clarify the dependence of $T^*$
on the parameters of the model.
This is also one of future tasks.

Finally, we briefly discuss a possibility to observe the present QCP
in actual materials.
If there exist Pr and/or Nd compounds in which the two-channel Kondo
effect is observed,
we propose to apply a high pressure to increase the hybridization $V$.
When the pressure is increased, as expected from Fig.~3(b),
we expect to observe quantum critical behavior in physical quantities
at a critical pressure, different from that in the two-channel Kondo phase.
At present, we cannot point out specific compounds,
but we expect that Pr 1-2-20 and Nd 1-2-20 compounds
may be candidates.

In summary, we have investigated the seven-orbital Anderson model
hybridized with three conduction bands by using the NRG technique.
We have observed the QCP characterized by $\log \phi$
between two-channel Kondo and Fermi-liquid phases,
when we have controlled $n$, $V$, and CEF parameters.
We hope that quantum critical behavior at this QCP can be
found in cubic Pr and Nd compounds in which two-channel Kondo effect
is observed.

\section*{Acknowledgment}

The author thanks K. Hattori, K. Kubo, and K. Ueda for discussions
on the two-channel Kondo phenomena.
This work has been supported by JSPS KAKENHI Grant Number 16H04017.
The computation in this work was partly done using the facilities of the
Supercomputer Center of Institute for Solid State Physics, University of Tokyo.

\end{document}